%% file: main.tex
\documentclass[runningheads]{llncs}
\usepackage[T1]{fontenc}
\usepackage{enumitem}
\usepackage{graphicx}
\graphicspath{{figures}}
\usepackage{booktabs}
\usepackage{adjustbox}
\usepackage{multirow}
\usepackage{comment}
\usepackage{caption}
\usepackage{subcaption}
\usepackage{hyperref}

\usepackage{color}

\begin{document}
\title{Boosting the Performance of Object Tracking with a Half-Precision Particle Filter on GPU\thanks{Preprint submitted for publication.}}
\titlerunning{A Half-Precision Particle Filter on GPU}

% Commented for the double-blind version
\author{Gabin Schieffer\inst{1} \and
Nattawat Pornthisan \inst{2} \and
Daniel Medeiros\inst{1} \and \\
Stefano Markidis \inst{1}\and
Jacob Wahlgren\inst{1} \and
Ivy Peng \inst{1}\
}
\authorrunning{G. Schieffer et al.}
\institute{KTH Royal Institute of Technology, Stockholm, Sweden\\
\email{ \{gabins, dadm, markidis, jacobwah, ivybopeng\}@kth.se}
 \and
Chulalongkorn University, Bangkok, Thailand \\
\email{w.winnaries@gmail.com}
}

\maketitle              % typeset the header of the contribution
\begin{abstract}
High-performance GPU-accelerated particle filter methods are critical for object detection applications, ranging from autonomous driving, robot localization, to time-series prediction. In this work, we investigate the design, development and optimization of particle-filter using half-precision on CUDA cores and compare their performance and accuracy with single- and double-precision baselines on Nvidia V100, A100, A40 and T4 GPUs. To mitigate numerical instability and precision losses, we introduce algorithmic changes in the particle filters. Using half-precision leads to a performance improvement of 1.5-2 $\times$ and 2.5-4.6 $\times$ with respect to single- and double-precision baselines respectively, at the cost of a relatively small loss of accuracy.

\keywords{Particle Filter \and Half-Precision \and Reduced Precision \and GPUs}
\end{abstract}

\input{introduction}
\input{background}
\input{method}

\input{result}

\input{conclusion}

 \bibliographystyle{splncs04}
 \bibliography{main}

\end{document}

%% file: introduction.tex
\section{Introduction}
The particle filter method is a critical algorithm for enabling automatic object or video tracking, e.g., automatically locating one or more moving objects over time using a camera. Today, it is widely used to support and improve autonomous driving and in a wide range of other applications, including video surveillance, sensor networks, signal processing, robot localization, and time-series forecasting~\cite{parallelpf,hsiao2005particle,jaward2006multiple}. For its central role in developing emerging technologies, such as autonomous driving, it is crucial to design, and develop high-performance, accelerated, yet accurate, particle filters that can provide real-time or near real-time object tracking capabilities. This research investigates the design and development of accurate particle filters with Graphical Processing Units~(GPU) and half-precision data. Ideally, the use of half-precision calculations on CUDA cores could lead to double the performance of the operations in single-precision on Nvidia GPUs and this work studies the challenges, achievable performance and accuracy on half-precision particle filters.

At this heart, the particle filter technique, also known as Sequential Monte Carlo~(SMC) method~\cite{gordon1993novel}, uses random sampling to simulate complex systems, such as object movement in a real environment.  Particle filters are particularly powerful in cases where the underlying system evolves over time and new observations, potentially affected by an error, become available sequentially, such as in a video stream. Intuitively, the fundamental idea of the particle filter is to approximate the underlying probability density function using a weighted set of samples (the so-called \textit{particles}). At each time step, particles are propagated using the system's transition model, and their weights are updated based on their likelihood to agree with the measurement from the system. After the update, particles with low weights are discarded, and only particles with high weights are re-sampled to create offspring particles for the next time step. Particle filter algorithms are compute-intensive as the propagation, weighting, and resampling calculate for each particle at every time step without significant data movement or synchronization. Previous works of parallelizing particle filters onto parallel computing, e.g., multi-core processors and GPUs~\cite{hendeby2010particle,chitchian2013adapting,parallelpf}, have achieved significant speedup. Recently, GPUs have provided increasing hardware support for low-precision arithmetic operations, motivated by machine learning workloads that are often compute-intensive but resilient to precision loss. In this work, we explore the hardware support for half-precision operations (FP16) on GPU to understand their impact on the performance and accuracy of particle filter algorithms. Differently from previous works~\cite{ho2017exploiting,markidis2018nvidia,haidar2018harnessing}, we focus specifically on pure FP16 operations and CUDA cores instead of mixed-precision FP16-FP32 calculations and Tensor Cores.

The main contributions of this work are the following:
\begin{itemize}[noitemsep,leftmargin=*,topsep=0pt]
    \item We developed an optimized half-precision particle filter for an object-tracking application on GPUs and associated double- and single-precision baseline implementations.
    \item We analyzed the numerical instability, performance, pipeline utilization, and accuracy of double-, single-, and two half-precision implementations on Nvidia A100, V100, A40, and Tesla T4 GPUs.
    \item We identified the performance bottlenecks in the baseline half-precision version and achieved $2-6\times$ speedup in the optimized version
    \item We characterized the impact of the number of threads per block on the particle filter and achieved a further 1.5-1.75$\times$ speedup on A100 and V100.
\end{itemize}

%% file: background.tex
\section{Background}\label{sec:bg}
%\subsection{Particle Filters on GPU}
The fundamental idea of the particle filter is to approximate the underlying probability density function $p$ using a set of weighted particles at each time step $t$ by 
$$p(\mathbf z_{1:t} \,|\, \mathbf y_{1:t}) = \sum_{k=0}^K \tilde w^k_t \delta(\mathbf z_{1:t} - \mathbf z_{1:t}^k)$$ 
where $y_t$ is a stochastic observation of the hidden state $z_t$ and $\tilde w^k_t$ is the corresponding normalized weights at time $t$ for $k$-th particle \cite{schon2010solving}. in concrete terms, for object tracking videos, the hidden state $z_t$ can be defined as the actual coordinate of the object in the image after $t$ frame, while $y_t$ is either the color histogram of the pixel in RGB image or the light intensity in monochrome image~\cite{jaward2006multiple}.

Particle filter algorithms are computationally intensive as they must iterate all particles at each time step $t$. Therefore, GPU is a natural fit for acceleration. Each time step generally goes through three main stages -- particle propagation, weighting, and resampling~\cite{hendeby2010particle,murray2016parallel}. After the initialization, the propagation algorithm creates particle samples from a set of given ancestors using a predefined transition model of the system. Then, the generated samples are incorporated with the observation $y_t$ to recalculate and normalize the particles' weights so that particles with low likelihood to the observation have reduced weights and discarded. Finally, the resampling step selects a set of important particles, i.e., high weight and high likelihood, to generate their corresponding number of offsprings for the next time step. Resampling is a critical step to focus the computational cost on those important samples that are in good alignment with measurements. 

Parallelizing the initialization, propagation, and weighting steps is straightforward as they are naturally data parallel. However, resampling requires cumulative operations on all particles, introducing synchronization among threads~\cite{parallelpf}. Therefore, resampling is the dominant stage in particle filters at a large number of particles.

%\subsection{Reduced-Precision Operations on GPUs}
The IEEE 754-2008 standard defines double, single, and half precision as a binary floating-point number format that occupies 64, 32, and 16 bits, respectively. While double precision, i.e., FP64, is commonly used in scientific computing, popular emerging workloads like neural networks and image processing have achieved significant performance boost from half-precision operations while resilient to precision loss.
In this work, we explore half-precision (FP16) on CUDA cores. The IEEE 754 defines that FP16 format has 1 sign bit, 5 exponent bits, and 11  significand bits (10 bits stored), achieving a value range of $\pm65,504$ and $\log_{10}^{2^{11}} \approx 3.311$ decimal digits.

CUDA provides two FP16 data types, i.e., \verb|half| and \verb|half2|. The former is a scalar data type and the latter is a vector of two elements. Converting a \verb|double| type to \verb|half| type in C++ is straightforward, except that several mathematical functions such as square root and exponential require a special half-precision replacement. Theoretically, the throughput of FP16 operations is twice FP32's throughput. However, on CUDA Cores, hardware arithmetic instructions operate on two 16-bit floating points at a time \cite{nvidiamp}. Therefore, for high utilization of hardware, operations on two \verb|half| values must be combined into a single operation on \verb|half2|. 
%Thus, naively converting data types from FP32 to FP16 (i.e., \verb|half| data type in CUDA) would not increase performance and may even degrade performance, as shown in our evaluation of the baseline half-precision version in Section~\ref{sec:opt}.

\section{Related Work}
The first implementation of a particle filter on GPU was proposed in~\cite{hendeby2010particle}, which explores parallelization opportunities in the classic particle filter, and propose a GPU-accelerated version. %In this work, the resampling step of the particle filter algorithm is listed as the most challenging step to parallelize, because of the cumulative summation performed during this step.
The Metropolis and rejection resampling algorithms were evaluated in \cite{murray2016parallel} to avoid the global reduction operation required in standard multinomial, stratified, and systematic resampling algorithms. The authors also evaluated numerical stability and bias in single precision. An approach to further improve the resampling step of the particle filter implementation on GPU was proposed in \cite{parallelresampling}. The authors evaluate their approach against both the Metropolis method and rejection resampling.

% Reduced precision on GPU
Modern GPU architectures offer tremendous performance in mixed-precision arithmetic. Common challenges of converting a CUDA program from single-precision to half-precision are detailed in~\cite{ho2017exploiting}. In particular, their findings indicate that the use of the \verb|half2| datatype, which combines two half-precision numbers in one structure, is crucial for performance. We use this technique in our work when converting the particle filter code to use half-precision number representation.
In addition, precision loss is an important source of concern when using half-precision. This issue is studied in \cite{markidis2018nvidia}, where the authors provide an analysis of Tensor Cores performance for large matrix multiplication in mixed-precision, and proposes a refinement technique to decrease precision loss. In \cite{haidar2018harnessing}, the authors propose to use an iterative refinement algorithm to improve the solution of their low-precision matrix factorization approach on Tensor Cores. 

%% file: method.tex
\section{Design, Implementation \& Optimization}
\begin{figure}[bt]
    \centering
    \includegraphics[width=0.9\textwidth]{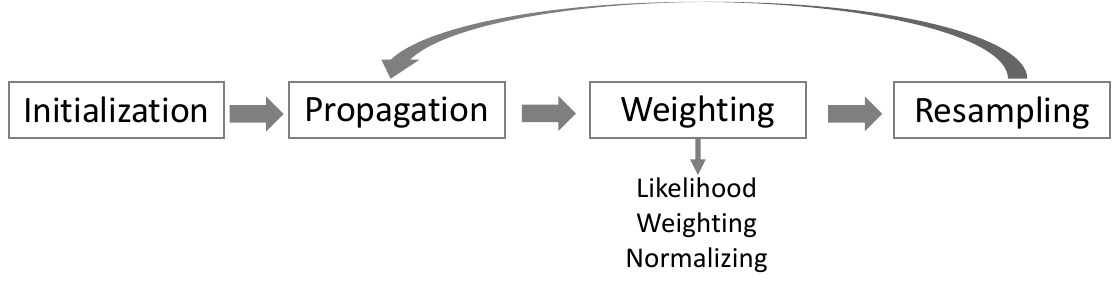}
    \caption{After the initialization, in each timeframe, the particle filter algorithm iterates over three main phases: propagation, weighting, and resampling.}
    \label{fig:workflow}
\end{figure}

We design and optimize a particle filter object tracking application to evaluate the effect of half-precision data types on Nvidia GPUs. The application tracks a moving object in a two-dimensional video. The baseline implementation follows the same algorithmic structure as in the Rodinia particle filter~\cite{che2009rodinia}, but has been improved on the overall organization and extended with single- and half-precision specialization. Fig.~\ref{fig:workflow} presents the main workflow, where after the initialization phase, three stages are repeated in each time frame, i.e., an iteration consisting of six kernels -- propagation, likelihood, maximum finding, weighting, normalizing, and resampling. For most kernels, their half-precision implementation is algorithmically different from their double- and single-precision counterparts because they have to work with the vector type to improve hardware utilization.

The propagation kernel takes a selected set of ancestors as the input.
%or merely an array of XY-coordinates of initial particles at the first time frame.
Propagation means that the next generation of each particle is sampled according to the true distribution of transition conditional density (denoted as $p(\mathbf{z}_{t+1} | \mathbf{z}_t)$), which is a property given on the specific system. Specifically,
\begin{equation} 
p(x_t^k \,|\, x_{t-1}^{a_k}) = \mathcal N\left( 1.0 + x_{t-1}^{a_k},  5.0^2 \right)
\end{equation}
\begin{equation} 
        p(y_t^k \,|\, y_{t-1}^{a_k}) = \mathcal N\left(
                2.0 + y_{t-1}^{a_k},
                2.0^2 
        \right) 
 \end{equation} 
 where $k$ is the particle index and $a_k$ is an ancestor index of particle $k$ returned from the resampling kernel in the last time frame. The next generation of particles are sampled from double-precision space, then converted to a target precision if reduced-precision is used because cuRAND is used for the random number generation on GPUs and does not support half-precision value. 

The likelihood kernel calculates the log-likelihood of observing the intensity around each particle at time $t$ by 
\begin{equation} \label{eq:likelihood}
   log^{likelihood} = L^k = \frac{(I_{ijk} - 100)^2 - (I_{ijk} - 228)^2}{50 \times N},
 \end{equation} 
where the two constants, 100 and 228, represent the mean intensity of the background and foreground, while $I_{ijk}$ refer to the pixel intensity at $i$th row and $j$th column at frame $k$ and $N$ refer to a total number of the pixel value to be calculated per particle. The likelihood kernel is parallelized pixel-wise across threads. For example, if the number of particles is set at 512 and there are 10 points around the particle to be evaluated, the kernel will launch a total of $512 \times 10 =$ 5,120 tasks. Note that the number of tasks is reduced by half on half-precision implementation, as each thread takes care of two pixels simultaneously. 

Each particle's weight is re-calculated and normalized at time $t$. Firstly, the max-finding kernel finds the highest likelihood among the particles by
\begin{equation} 
    L^k = \left(\frac{I_{ijk} - 100}{\sqrt{50 \times N}}\right)^2 - \left(\frac{I_{ijk} - 228}{\sqrt{50 \times N}}\right)^2. 
 \end{equation} 
Next, the weighting kernel calculates the unnormalized weight by 
\begin{equation} 
    w_{t}^k = w_{t-1}^k \exp{\{L^k - \max_i{L^i}\}},
 \end{equation} 
and performs a sum-reduction operation on weights. Then, given the pre-calculated sum, the normalizing kernel normalizes the weights and prepares the cumulative distribution, which is essentially an inclusive-prefix-sum \cite{gpugem3}, for the next stage. 

Resampling selects a set of particles to become the ancestors for the next generation of particles based on their weights and the resampling algorithm. The most straightforward resampling scheme would be multinomial resampling, which is just a categorical distribution with the probability of choosing $k$th particle as an ancestor equal to its normalized weight. In our case, to be consistent with the original Rodinia implementation, we employ a \textit{systematic resampling} algorithm that divides the normalized space into $N$ partitions, where $N$ is the number of particles and takes one sample from each partition \cite{liresampling}. The parallelization scheme among threads is similar to the other kernels, i.e., in single- and double-precision each thread finds the ancestor for one particle, while in half-precision, each thread finds the ancestors for two particles at a time.\\

\noindent \textbf{Implementation}. For the ease of running experiments, we utilized C++ templates to allow multi-precision support for the kernel. Switching between data types is as effortless as changing a function call, e.g., from \verb|particleFilter<double>| to \verb|particleFilter<half>(...)|. Note that single and double precision use the same kernel implementations, where each thread process one particle, while half-precision kernels change to \verb|half2| datatype and use a different implementation where each thread processes two particles packed into one \verb|half2| element.\\ %The implementations presented in this work are available online on GitHub \footnote{\url{https://github.com/Winnaries/mixed-precision-bpf}}.

\noindent \textbf{Numerical Stability Optimization.} First, we analyze the numerical stability in reduced precision and propose our optimizations. In Propagation, the algorithm operates on a low range of values, and thus, there is no significant numerical instability to be concerned. The likelihood kernel can have a numerical overflow when the square operations in Eq.~\ref{eq:likelihood} result in large values. Instead of computing Eq.~\ref{eq:likelihood} directly, we move the denominator to the square operator too to avoid the problem. In the normalizing kernel, since the exponents can grow arbitrarily large and lead to a particle with infinite weight, the resulted normalized weight can vanish. One way to solve the problem is to employ the log-sum-exp, used in statistical modeling and machine learning. Nevertheless, it requires one more reduction operation to find the maximum likelihood value. On the other hand, the Gumbel-max scheme allows the algorithm to stay on a log space and draw samples from the unnormalized log weight instead, limiting the applicable choice of resampling algorithms to be multinomial resampling only. Given these trade-offs, we decide to use a log-sum-exp scheme to maintain numerical stability and consistency with the existing implementation. For parallelization on GPU, each thread processes each particle in the normalizing step and they jointly perform a reduction operation through a sequential addressing technique proposed by~\cite{seqreduction}. Half-precision version is similar except that each thread processes two particles at a time. The resampling kernel only involves a series of conditional checks. % there is no numerical instability to be concerned with when using a low-precision floating point. 

%\subsection{Optimizations for Computing Efficiency}
\begin{figure}[ht]
    \centering
    \includegraphics[height=2.5cm]{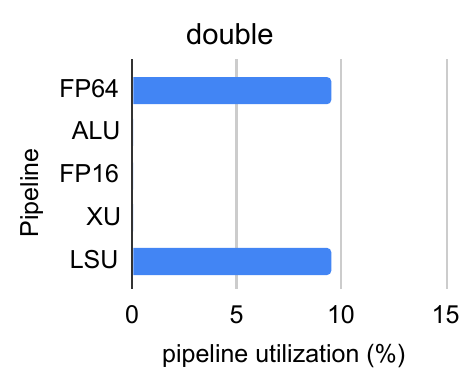}\hfill
    \includegraphics[height=2.5cm]{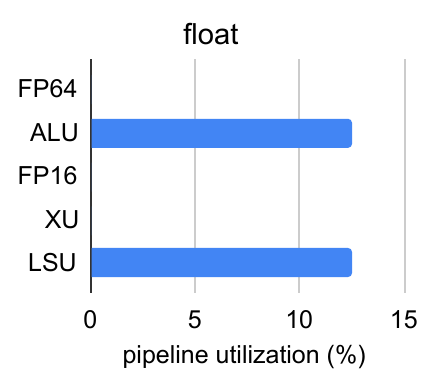}\hfill
    \includegraphics[height=2.5cm]{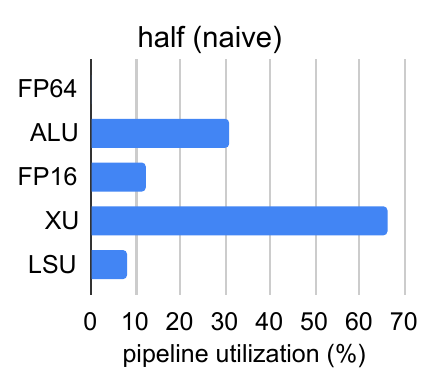}\hfill
    \includegraphics[height=2.5cm]{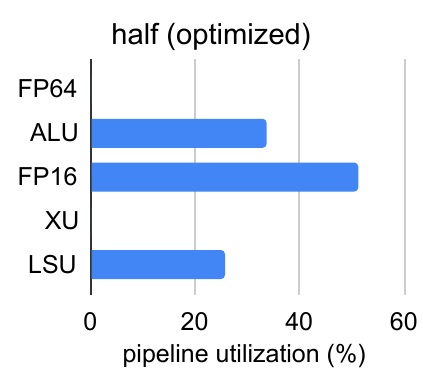}
    \caption{The pipeline utilization for double-, single-, half-precision implementations on V100.}
    \label{fig:pipeline}
\end{figure}
\noindent \textbf{Performance Optimization.} Next, we describe optimizations on hardware pipeline utilization on GPU. The initial implementation of half-precision implementation runs slower than single-precision implementation in our tests, even though it is already processing two packed particles in each \verb|half2| element. Fig.~\ref{fig:pipeline} presents the profiling results of pipeline utilization obtained from the NVIDIA Nsight Compute. After analysing the instruction mix and hardware pipeline, we find that FP16 pipeline is not as highly utilized as other pipelines like ALU and XU. The utilization of XU pipeline is surprisingly high, reaching 66\%. The XU pipeline is responsible for special functions like sin, cos, and reciprocal square root and also int-to-float and float-to-int type conversions. The utilization of ALU pipeline is as expected as conversion between FP32 and FP16 may be included, e.g., NVIDIA Ampere architecture. Also, not all functions in CUDA library supports half-precision, e.g., cuRAND API, and single-precision operations are still used. We optimized the resampling kernel based on these findings, by reducing the computation of reciprocal operations and data casting operations for particles with a saved constant. The optimized half-precision implementation not only removes the bottleneck on XU pipeline but also much improved FP16 utilization, increasing from 10\% to 50\% as shown in Fig.~\ref{fig:pipeline}.

%% file: result.tex
\section{Experimental Setup \& Evaluation}\label{sec:eval}
We evaluate our particle filter implementation for three different precisions,  namely \verb|double|, \verb|float| and \verb|half| on four different Nvidia GPU testbeds. We summarize the hardware and software environments in Table~\ref{tab:setup}. For reproducible comparison across different precision implementations, the same random number generator seed is used.
\vspace{-20pt}
\begin{table}[ht]
\caption{Nvidia GPUs hardware and software testbeds.\label{tab:setup}}
\begin{adjustbox}{width=0.9\linewidth,center}{
    \centering
    \begin{tabular}{|c|c|c|c|c|}
    \hline
    \textbf{GPU} &\textbf{Compute Capability} &\textbf{CUDA Version} &\textbf{GPU Memory} &\textbf{Peak FP32 (non-Tensor)}\\\hline
    Nvidia V100 &7.0 &11.7 &32GB HBM2  & 14.0 TFLOPS\\\hline
    Nvidia A100 &8.0 &11.7 &40GB HBM2e & 19.5 TFLOPS\\\hline
    Nvidia A40  &8.6 &12.0 &48GB GDDR6 & 37.4 TFLOPS\\\hline
    Nvidia T4   &7.5 &11.7 &16GB GDDR6 & 8.1  TFLOPS\\\hline
    \end{tabular}
}\end{adjustbox}
\end{table}

\begin{figure}[t]
  \centering
      \begin{minipage}[c]{0.3\linewidth}
        \centering
        \includegraphics[width=.9\textwidth]{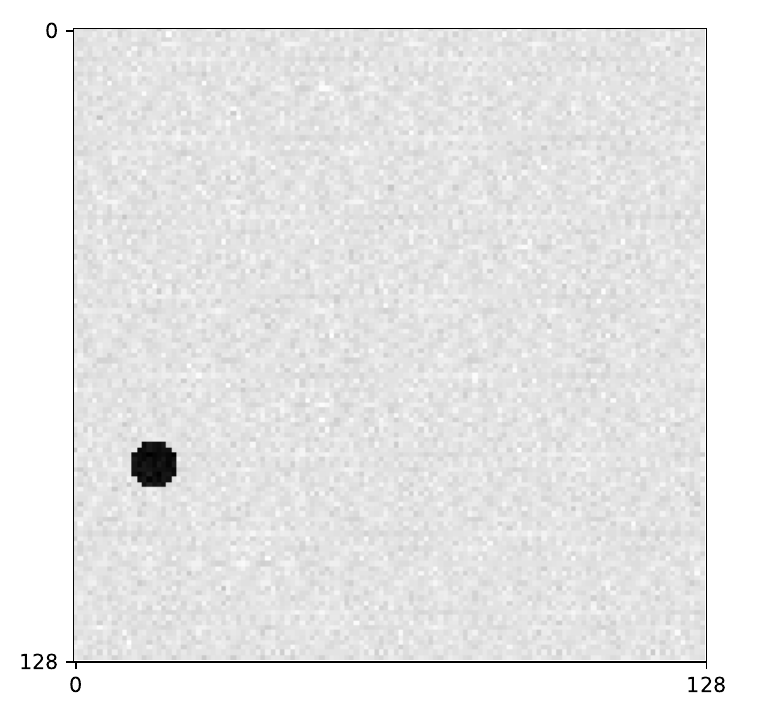}
        \caption{Object tracking in a $128\times128$ video for evaluation.}
    \end{minipage}\hfill
    \begin{minipage}[c]{0.6\linewidth}
    \centering
   \includegraphics[width=\textwidth]{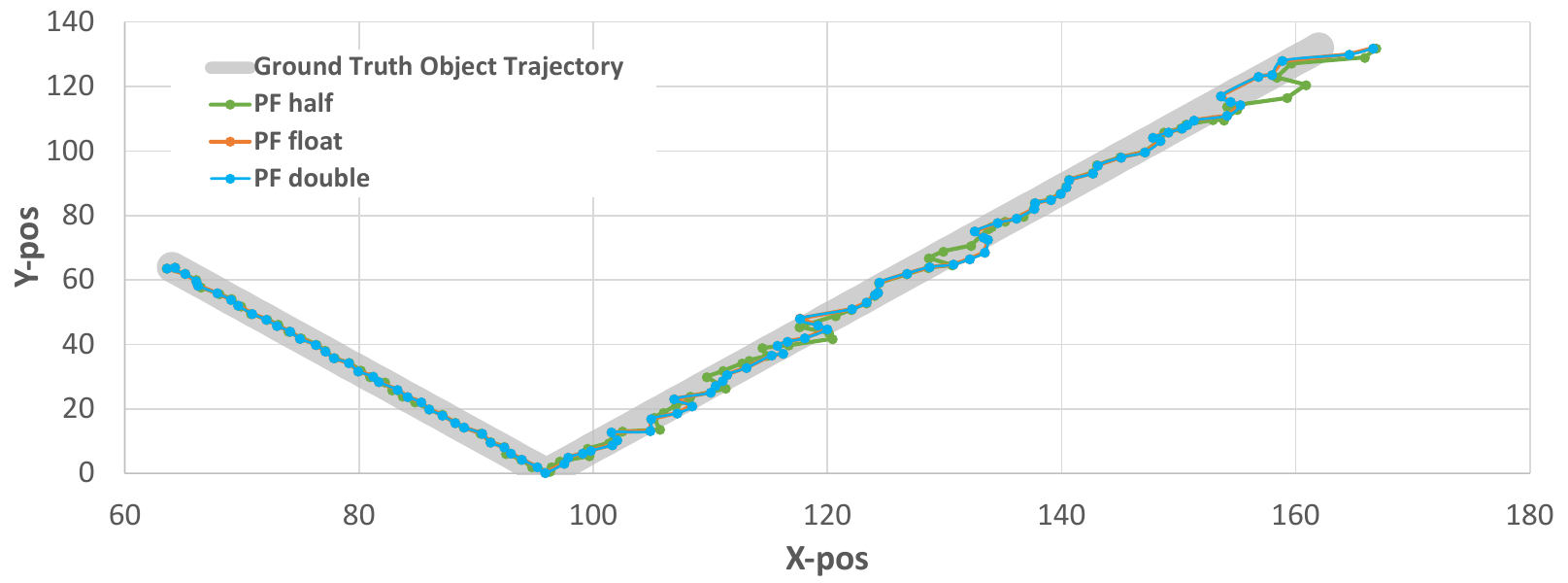} % accuracy_128p_100frames.pdf
   \caption{Object tracking using full, single, and half precision particle filter implementations and comparison with the ground truth.}
   \label{fig:accuracy}
   \end{minipage}
\end{figure}

\noindent \textbf{Verification \& Accuracy Results.} As the first step of this work, we verify the correctness of our implementations and assess the impact of performing object tracking with lower precision. To achieve this, we execute the particle filter algorithm on a 100-frame video with a resolution of $512\times512$ pixels, as proposed in the original \texttt{Rodinia} particle filter mini-application. This synthetic video represents a circular-shaped object moving in a two dimensional plane $xy$. The object moves towards the wall at $y=0$ and bounce back specularly. A single level of pixel intensity is chosen for the background and the foreground. Finally, a Gaussian noise is added to the frame. An example of frame and the tracked trajectories produced by the particle filter with 128 particles and the ground truth trajectory are presented in Fig.~\ref{fig:accuracy}.

We observe that the baseline double-precision implementation follows accurately the ground truth trajectory. In addition, the predictions obtained with the single-precision implementation exactly match those obtained with the baseline implementation, for all input frames. This first observation indicates that the reduction of precision from double-precision to single-precision did not have any impact on the accuracy of the predictions. The predictions obtained with our half-precision implementation provide comparable accuracy with regard to the ground truth. These results validate the use of lower-precision number representations in the particle filter algorithm, as both single- and half-precision implementations provided accurate tracking of the object in this experiment, and exhibited similar results compared to the baseline double-precision implementation. The results on accuracy confirm that particle filter algorithm and also statistical learning, in general, are resistant to low-precision computation. %\cite{chrisPRforDL, relmPRforDL, nvidiamp}
% Interestingly, in this experiment, the half-precision implementation provides better predictions compared to the baseline implementation for a limited number of frames, when the tracked object is in the region $x \in \{125,135\}$.
% The results on accuracy confirm that particle filter algorithm and also statistical learning, in general, are resistant to low-precision computation \cite{chrisPRforDL, relmPRforDL, nvidiamp}. Unlike machine learning, approximate inference, including particle filters, is relatively small in terms of problem size. Although they still hugely benefit from GPU-enabled parallelism \cite{parallelpf}, reduced precision might not be as necessary unless highly-critical real-time features are needed, or a very large input size is present.

\begin{figure}
    \begin{minipage}[c]{0.49\linewidth}
        \centering
        \includegraphics[width=0.98\textwidth]{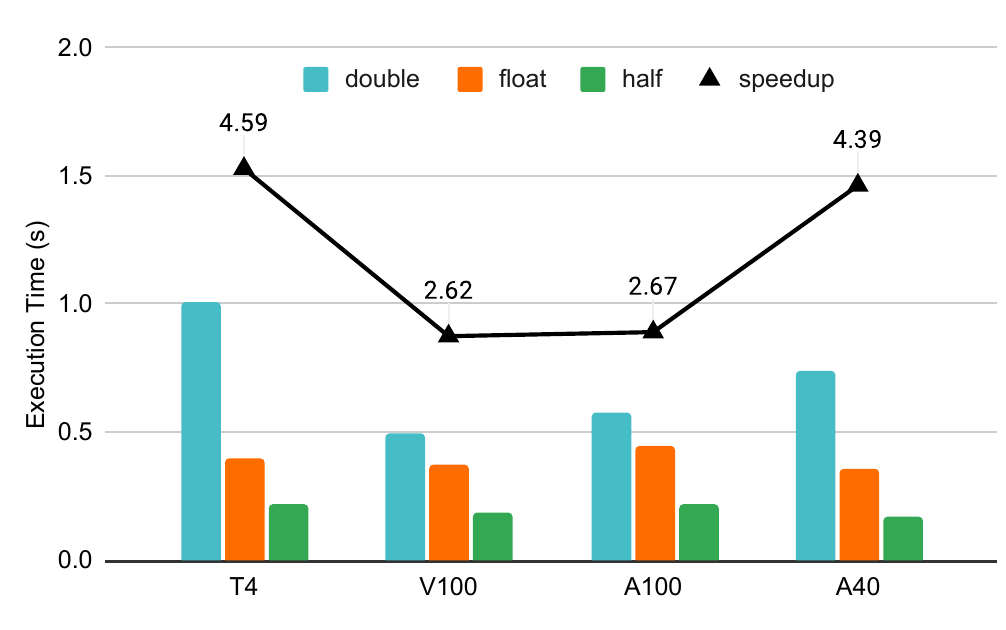} 
    \end{minipage}
    \hfill
    \begin{minipage}[c]{0.49\linewidth}
        \centering
        \includegraphics[width=0.98\textwidth]{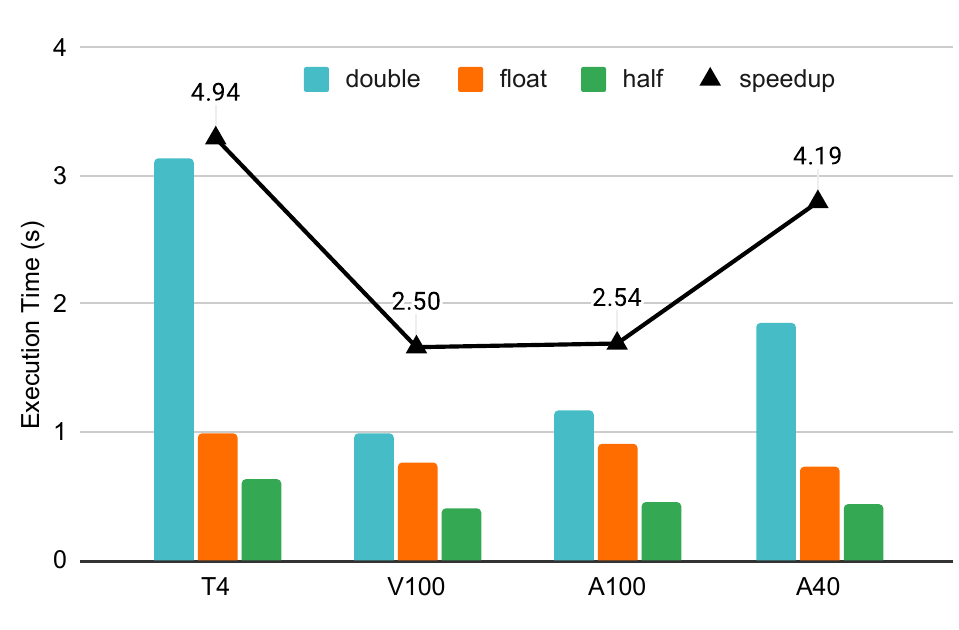}
    \end{minipage}
    \caption{Execution time and speedup for particle filters with 32k and 64k particles.}
     \label{fig:speedup}
\end{figure}

\noindent \textbf{Overall Performance.} As second step of our study, we evaluate the average execution time for 100 runs of the particle filter at three levels of precision and speed up achieved by using reduced precision. We perform this evaluation for two problem sizes by using either 32k or 64k particles and 128 threads per block. Fig.~\ref{fig:speedup} show the results in terms of execution time, along with the speedup provided by our optimized half-precision implementation, over the \verb|double| baseline. 
% Very general: for all situations, double > float > half
For both problem sizes, we consistently observe that the single and half precision implementations outperform the double-precision baseline, on all four GPU architectures. Furthermore, our half precision implementation outperforms the single precision implementation in all test cases.
% - 4.5x speedup for 4x memory footprint reduction -> higher throughput -> because of the use of half2 i think
The highest speedup value is reached on the T4 GPU, namely $4.59\times$ and $4.94\times$ for 32k and 64k particles, respectively. While the execution time for the double-precision implementation varies significantly across architectures, the execution time for both single- and half-precision implementations is consistent across all four GPU architectures. 
% - Two speedup values: ~2x (V100 and A100) and ~4x (T4 and A40), why?
We observe that the speedup provided by our half-precision implementation is above $4\times$ on both A40 and T4, while it is close to $2.5\times$ on both V100 and A100. This difference is induced by the low performance observed in double-precision for A40 and T4, compared to the two other GPUs. 
We explain this difference by the target workloads those GPUs were designed to run. The Tesla T4 GPU is designed primarily for machine learning workloads, which requires higher single-precision performance and relatively low double-precision performance. A40 also follows the same trend, where single- and half-precision performance are favored in GPU design, producing relatively low double-precision performance.

% \todo{Tesla T4 has NVIDIA’s “Turing” architecture, which includes Tensor Cores and CUDA cores (weighted towards single-precision).}

\begin{figure}
\begin{minipage}[t]{0.47\linewidth}
    \centering
    \includegraphics[height=2.3cm]{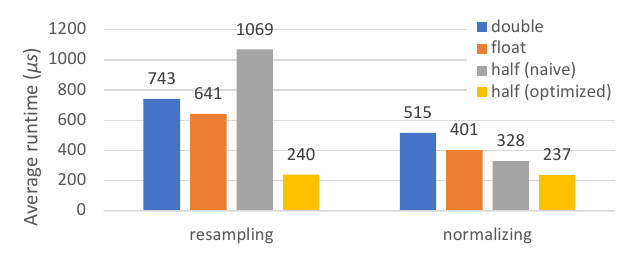}
    \caption{Average runtime for the resampling and normalizing kernels, for the three levels of precision. Both the naive and optimized half-precision implementations are shown.}
    \label{fig:runtime_kernel}
\end{minipage}
\hfill
\begin{minipage}[t]{0.47\linewidth}
    \centering
    \includegraphics[width=\textwidth]{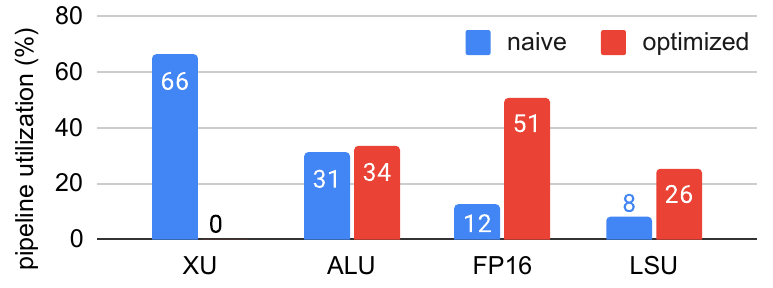}
    \caption{Pipeline utilization for the resampling kernel, for the naive and optimized half-precision implementations, on V100. Pipelines with utilization below 5\% are not shown.}
    \label{fig:pipeline_results}
\end{minipage}
\end{figure}

\noindent \textbf{Performance Breakdown.} Next, we investigate in detail the effect of using lower levels of precision on the execution time of essential kernels. We also include a comparison of the optimized and naive versions of our half-precision implementation, both in terms of kernel runtime, and GPU pipeline utilization.
The profiling results of the two major kernels -- resampling, and normalizing -- are presented in Fig.~\ref{fig:runtime_kernel}. They were obtained through profiling using NVIDIA Nsight Systems, with 8192 particles, on V100 GPU. In this experiment 100 samples for each kernel are captured, as we run the particle filter on a 100-frame video. Our results show that the introduction of single precision improved the runtime compared to the double-precision baseline, for both the resampling and normalizing kernels, by respectively 16\% and 28\%. This observation is coherent with the overall runtime improvement observed previously.

The naive version of the half-precision implementation does not provide satisfactory runtime improvement for the resampling kernel. However, the optimized half-precision implementation provided a significant speedup over both the double-precision and single-precision implementations for the resampling kernel, namely $\times3.0$ and $\times2.7$, respectively. These observations validate the necessity of our optimization efforts. %, described in Section~\ref{sec:optimization_efforts}.

%\begin{figure}
%    \centering
%    \includegraphics[height=2.3cm]{kernel_runtime_8192p.pdf}
%    \caption{Average runtime for the resampling and normalizing kernels, for the three levels of precision. Both the naive and optimized half-precision implementations are shown.}
%    \label{fig:runtime_kernel}
%\end{figure}

%\begin{table}[bt]
%\caption{Average runtime for the resampling and normalizing kernels, for the three levels of precision. Both the naive and optimized half-precision implementations are shown.}\label{tab:runtime_kernel}
%\begin{adjustbox}{width=0.9\linewidth,center}{
%\centering\begin{tabular}{|c|c|c|c|c|c|}
%    \hline
%    \multirow{2}{*}{\textbf{Kernel}} &\multicolumn{4}{c|}{\textbf{Avg. Time ($\mu$s)} }\\
%     &Double Precision & Single Precision & Mixed Precision (naive) &Mixed Precision (optimized)\\\hline
%    resampling  & 743 & 641 & 1069 & 240 \\\hline
%    normalizing & 515 & 401 & 328 & 237\\\hline
%\end{tabular}}\end{adjustbox}
%\end{table}

To evaluate the effect of the optimizations on the device utilization, we profile the resampling kernel using NVIDIA Nsight Compute to collect statistics on the utilization of computing pipelines. The results for both naive and optimized half-precision implementations are detailed in Fig.~\ref{fig:pipeline_results}. 
We observe that the utilization of ALU, FP16, and LSU pipelines is significantly higher for the optimized version than for the naive one. In particular, the FP16 pipeline, which is responsible for arithmetic operations on half-precision numbers, exhibits a 12\% utilization in the non-optimized version, while the optimized version reaches a 51\% utilization of this pipeline. 
Additionally, we observe that the XU pipeline is being heavily utilized in the non-optimized version, while it is not utilized at all in the optimized version. This indicates precisely where our optimization had effect, and we make the hypothesis that this high utilization of the XU pipeline was limiting the performance, and explains the low FP16 utilization.\\
% Those observations show that our optimizations allowed to improve computation efficiency by increasing GPU utilization. 

% (Arithmetic and Logic Unit) is nearly fully utilized in the optimized implementation, while it only exhibits a 48\% utilization for the naive version. This difference is also noticeable for the FP16 pipeline, which performs arithmetic operations of half-precision numbers. This pipeline is only utilized at 8\% in the naive version, while it shows a 76\% utilization in the optimized version. 

%\begin{figure}[bt]
%    \centering
%    \includegraphics[width=.7\textwidth]{pipeline_utilization_barplot.pdf}
%    \caption{Pipeline utilization for the resampling kernel, for the naive and optimized half-precision implementations, on V100. Pipelines with utilization below 5\% are not shown.}
%    \label{fig:pipeline_results}
%\end{figure}

%\begin{figure}[ht]
%    \centering    \includegraphics[width=0.45\textwidth,height=70pt]{pipeline_util_half_old_full.jpg}
%    \includegraphics[width=0.45\textwidth,height=70pt]{pipeline_util_half_new_full.jpg}
%    \caption{The pipeline utilization of the naive (left) and optimized (right) implementations using half precision on V100.}
%    \label{fig:pipeline_results}
%\end{figure}
\begin{figure}[bt]
\begin{subfigure}[t]{0.50\textwidth}
    \centering
    \includegraphics[width=0.8\textwidth]{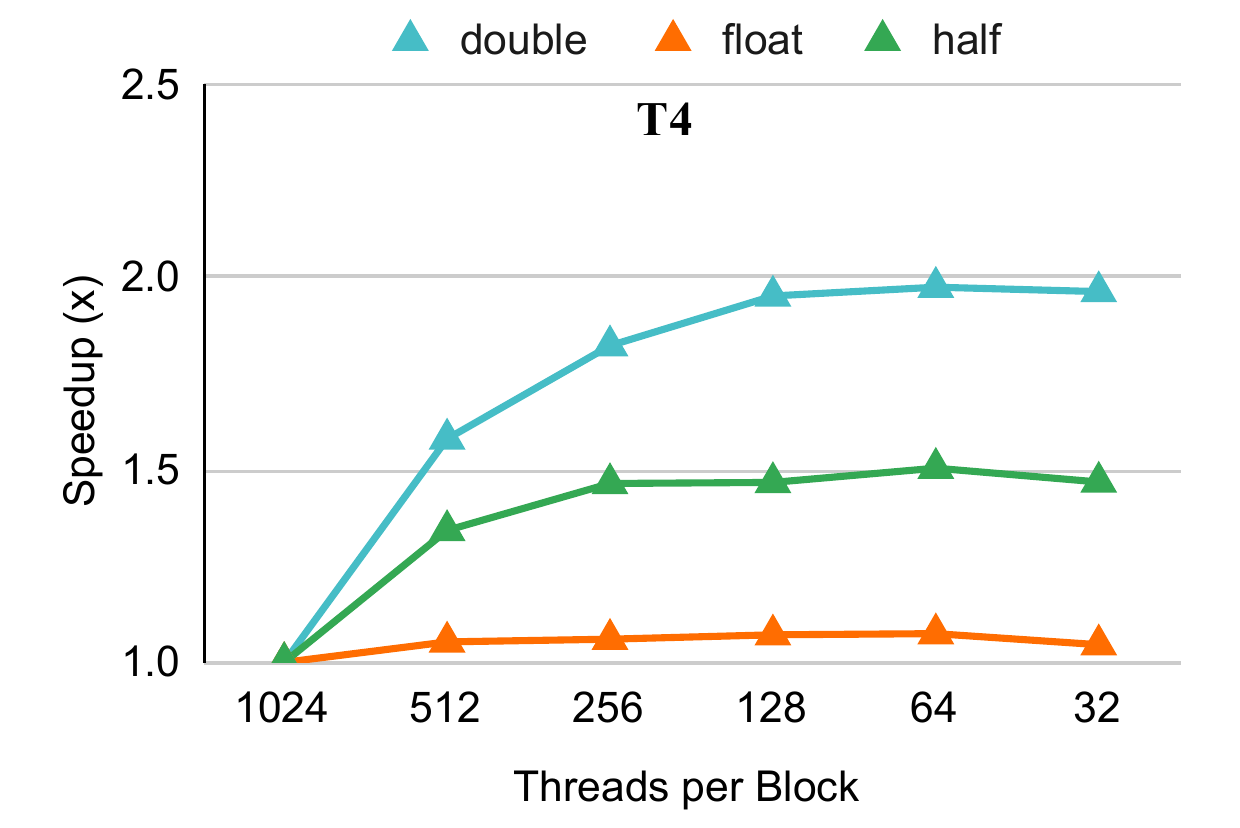}
\end{subfigure}\hspace{\fill} % maximize horizontal separation
\begin{subfigure}[t]{0.50\textwidth}
    \centering
    \includegraphics[width=0.8\linewidth]{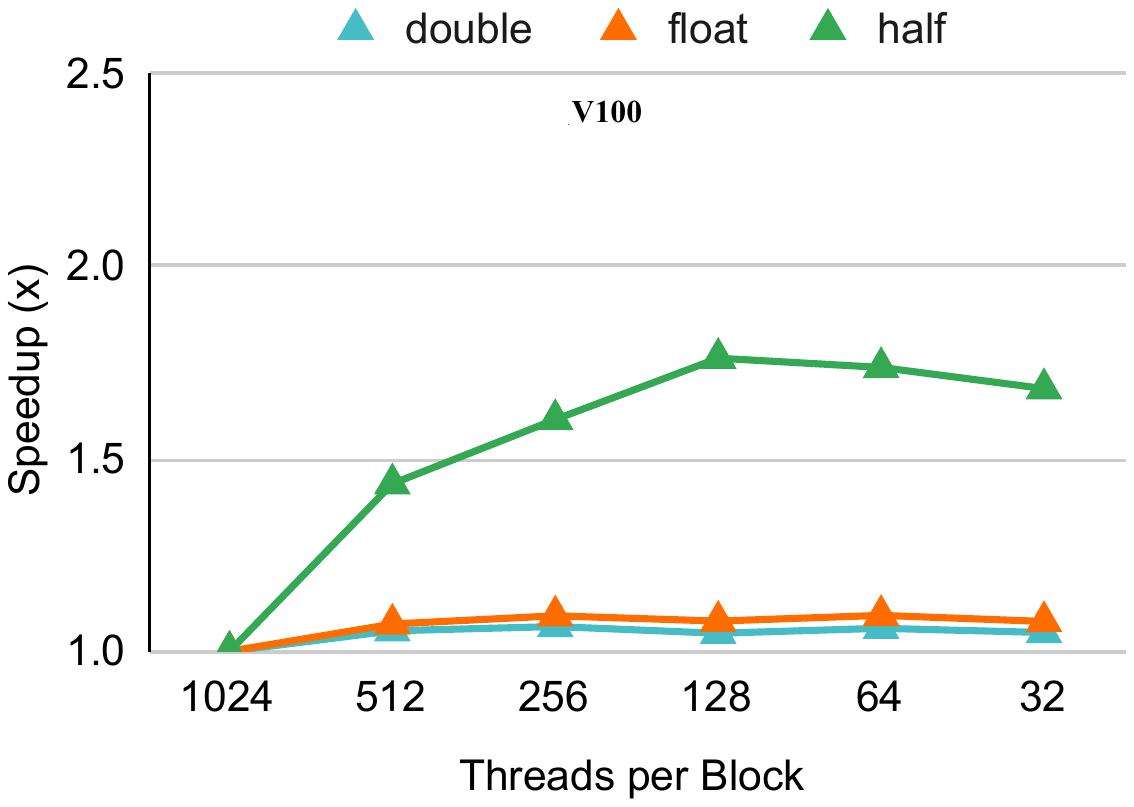}
\end{subfigure}
\begin{subfigure}[t]{0.50\textwidth}
    \centering
    \includegraphics[width=0.8\linewidth]{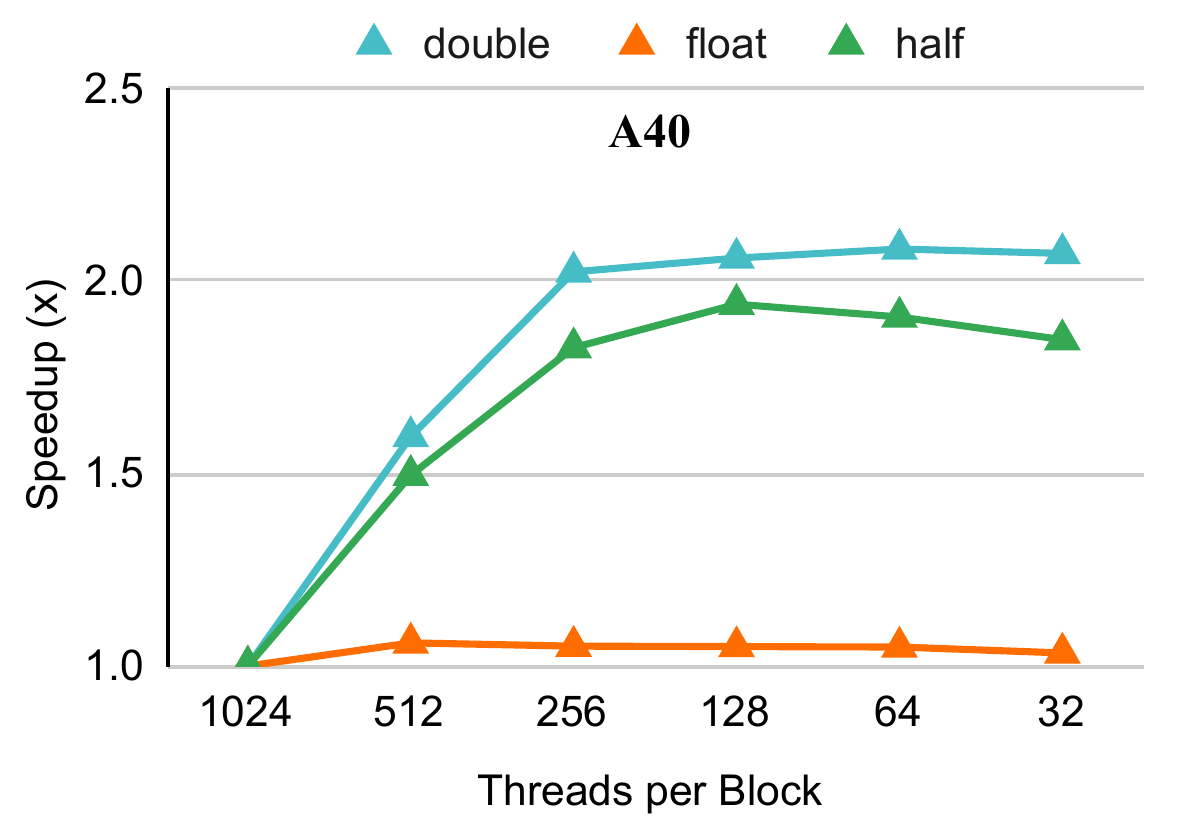}
\end{subfigure}\hspace{\fill} % maximize horizontal separation
\begin{subfigure}[t]{0.50\textwidth}
    \centering
    \includegraphics[width=0.8\linewidth]{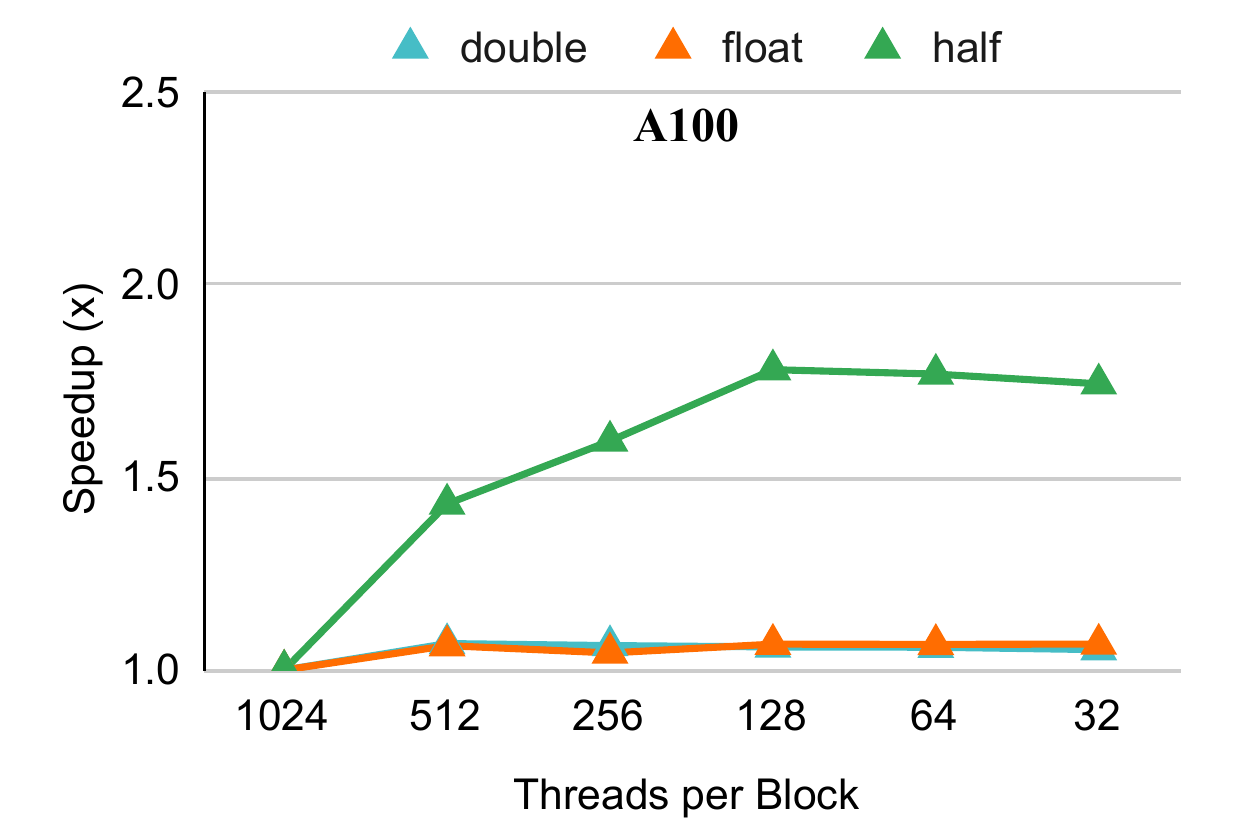}
\end{subfigure}
\caption{Performance impact of Threads per Block (TPB) on the three versions on four GPU architectures.}
\label{fig:tpb}
\end{figure}

\noindent \textbf{Impact of Thread Block Configuration.} Finally, we investigate the impact of thread block configuration for several GPU architectures. To perform this evaluation, we execute the three implementations using various numbers of threads per block, between 32 and 1024. The results in terms of speedup over the worst-performing configuration are shown on Fig.~\ref{fig:tpb} for the four GPU models. We observe that decreasing the block size below 1024 threads generally increases speedup up to 128 threads, where it starts to plateau.
Both V100 and A100 GPUs exhibit similar characteristics, where the speedup for double- and single-precision implementations is below $\times1.1$ for all block configurations. However, the performance of the half-precision implementation appears to significantly benefit from decreasing the number of threads per block. The maximum speedup for those two GPUs is $\times1.8$ when using a block size of 128.
For A40 and T4 GPUs, the double-precision version is the implementation which benefits the most from decreasing the block size. The maximum speedup of $\times2$ is reached when using 64 threads per block.
To a lower extent, the runtime of the half-precision implementation also improves when reducing the number of threads per block. As for the two other GPUs, the runtime for single-precision is hardly impacted by the block size. % \todo{why, add some explanation/hypothesis}

%% file: conclusion.tex
\section{Discussion and Conclusion}
In this work, we presented a multi-precision implementation of a particle filter on GPU, which supports half-, single-, and double-precision floating points. We showed that Using half-precision leads to a performance improvement of 1.5-2 $\times$ and 2.5-4.6 $\times$ with respect to single- and double-precision baselines respectively, at the cost of a relatively small loss of accuracy.

Our study shows that algorithm re-design is needed to effectively leverage half-precision operations in particle filter codes on CUDA cores. Our initial porting to half precision has already re-designed algorithms to use intrinsics to pack two FP16 values into one FP32 register using the vector type $half2$ for boosting the utilization of the FP32 pipeline. Still, it did not lead to performance benefits and even caused performance degradation due to the increased conversion instructions causing skewed utilization of hardware pipelines. Also, we note that re-designing large-scale applications to use intrinsics may not always be feasible in practice.

We evaluated the accuracy loss due to reduced precision in an object-tracking application. One limitation of our current design is that up to 64k particles can be used in a particle filter. This is a reasonable assumption because, unlike machine learning, approximate inferences, including particle filters, are relatively small in terms of problem size. Our work focuses on half-precision FP16 operations on CUDA cores, while previous works have shown mixed-precision on Tensor Core could bring even higher throughput but require refactoring algorithms into matrix-multiplication forms, which we plan to explore in future work.